\documentclass[11pt,a4paper]{emulateapj}
\usepackage{amsmath}
\usepackage{multirow}
\usepackage[export]{adjustbox}

\begin{document}
%%%%%%%%%%%%%%%%%%%%%%%%%%%%%
%%%%% Title of proposal %%%%% 
%%%%%%%%%%%%%%%%%%%%%%%%%%%%%

%\title{Deep Chandra Observations of nearby elliptical NGC~1404:

%Transport Phenomena and Microphysics in Hot Plasma}
\title{Deep Chandra observations of NGC~1404: cluster plasma physics revealed by an infalling early-type galaxy}

\author{Yuanyuan Su\altaffilmark{$\ddagger$1}}
\author{Ralph P. Kraft\altaffilmark{1}}
\author{Elke Roediger\altaffilmark{2}}
\author{Paul E. J. Nulsen\altaffilmark{1}}
\author{William R. Forman\altaffilmark{1}}
\author{Eugene Churazov\altaffilmark{3}}
\author{Scott W. Randall\altaffilmark{1}}
\author{Christine Jones\altaffilmark{1}}
\author{Marie E. Machacek\altaffilmark{1}}
\affil{$^1$Harvard-Smithsonian Center for Astrophysics, 60 Garden Street, Cambridge, MA 02138, USA}
\affil{$^2$E.A. Milne Centre for Astrophysics, Department of Physics and Mathematics, University of
Hull, Hull, HU6 7RX, United Kingdom}
\affil{$^3$Max Planck Institute for Astrophysics, Karl-Schwarzschild-Str. 1, 85741, Garching, Germany}

\altaffiltext{$\ddagger$}{Email: yuanyuan.su@cfa.harvard.edu}

%%%%%%%%%%%%%%%%%%%%%%%%%%%%%%%%%%%%%%%%%
%%%%% Body of science justification %%%%%
%%%%% and technical feasibility     %%%%%
%%%%%%%%%%%%%%%%%%%%%%%%%%%%%%%%%%%%%%%%%

%%
%% ENTER TEXT AND FIGURES BELOW
%%

\begin{abstract}

The intracluster medium (ICM), as a magnetized and highly ionized fluid, provides an ideal laboratory to study plasma physics under extreme conditions that cannot yet be achieved on Earth. NGC~1404 is a bright elliptical galaxy that is being gas stripped as it falls through the ICM of the Fornax Cluster. We use the new {\sl Chandra} X-ray observations of NGC 1404 to study ICM microphysics. The interstellar medium (ISM) of NGC 1404 is characterized by a sharp leading edge, 8 kpc from the galaxy center, and a short downstream gaseous tail. Contact discontinuities are resolved on unprecedented spatial scales ($0\farcs5=45$\,pc) due to the combination of the proximity of NGC 1404, the superb spatial resolution of {\sl Chandra}, and the very deep (670 ksec) exposure. At the leading edge, we observe sub-kpc scale eddies generated by Kelvin-Helmholtz instability and put an upper limit of 5\% Spitzer on the isotropic viscosity of the hot cluster plasma. We also observe mixing between the hot cluster gas and the cooler galaxy gas in the downstream stripped tail, which provides further evidence of a low viscosity plasma. The assumed ordered magnetic fields in the ICM ought to be smaller than 5\,$\mu$G to allow KHI to develop. The lack of evident magnetic draping layer just outside the contact edge is consistent with such an upper limit.
\end{abstract}

\section{\bf Introduction}

Contemporary X-ray observations, particularly {\sl Chandra} and {\sl XMM-Newton}, have revealed a wealth of substructure in the intracluster medium (ICM), even in relatively relaxed clusters. One important result is the ubiquitous presence of ``cold fronts," characterized by sharp surface brightness edges, resulting from contact discontinuities between gaseous regions of different entropies (see Markevitch \& Vikhlinin 2007 for a review).  
Cold fronts have two origins. 
First, they can be directly induced by dense and cool subgroups falling into a relatively diffuse and hot ICM (e.g., Abell~3667, Vikhlinin et al.\ 2001). 
Alternatively, off-axis mergers can disturb lower entropy gas at the bottom of a cluster potential well and 
bring it into contact with hotter ICM outside the cluster center (Ascasibar \& Markevitch 2006).
The former phenomenon is often referred as a ``merging cold front" and the
latter is called a ``sloshing cold front". 

The accretion of galaxies and subclusters not only impacts the ICM, it also affects the perturber.  
Often observed in nearby clusters are infalling galaxies with a truncated interstellar medium (ISM) (Su et al.\ 2015), a cold front at the leading edge, and a downstream tail of stripped galactic gas (M86--Forman et al.\ 1979; Randall et al.\ 2008; NGC~1400--Su et al.\ 2014; M49--Kraft et al.\ 2011). 

At temperatures of 1--10 keV, hot cluster gas is a highly ionized plasma. 
This makes the ICM an ideal plasma physics laboratory in which to study transport processes such as viscosity and thermal conductivity and to determine ill-constrained transport coefficients.  
ICM microphysics regulates cluster evolutionary processes, e.g., the dissipation of the kinetic energy injected by active galactic nuclei (AGN) (e.g., Fabian et al.\ 2005; Churazov et al.\ 2000), the thermalization and relaxation of the ICM from the cluster center to large radii.
The role of ICM magnetic field is considered to be small on a macroscopic scale since magnetic pressure is often insignificant compared to the thermal pressure in relatively quiescent medium, while it becomes non-negligible when it comes to the microscopic scale in that any non-zero magnetic field is able to alter the gyroradius of electrons (e.g., ZuHone \& Roediger 2016; Squire et al.\ 2016).    
X-ray observables provide excellent probes of the physical conditions in ICM plasma. For instance, surface brightness fluctuations in bright galaxy clusters were used to probe gas motions that are suggestive of a turbulent ICM (e.g., Churazov et al.\ 2012; Zhuravleva et al.\ 2015). Recently, a low level of turbulence was unveiled by the calorimeter on board {\sl Hitomi} (Hitomi Collaboration 2016).  

The sharp edges in temperature and surface brightness are not expected to be hydrodynamically stable. This is partly due to thermal conduction and diffusion, although even a very weakly magnetized plasma can suppress these processes.  
The front can also be distorted by Kelvin-Helmholtz instabilities (KHI) driven by  
shear flows around the extended low entropy substructures (Chandrasekhar 1961; Lamb 1932). They appear as ``horn" or ``roll" features in X-ray images of cold fronts (e.g., Roediger et al.\ 2015a).  
KHI coupled with turbulent mixing would accelerate ablation of low entropy regions by the ambient hot ICM. Viscosity, operating near the level expected when the particle mean free paths are determined by Coulomb collisions (Spitzer 1956), could prevent the growth of instabilities (Roediger et al.\ 2013). The presence of strong magnetic field would further suppress the mixing between gaseous regions of different entropies (Chandrasekhar 1961). 
It is worth noting that a non-zero intrinsic width of the cold front may also limit the growth of KHI at regions near the stagnation point (Churazov \& Inogamov 2004).

Deep {\sl Chandra} observations of the ICM are particularly suitable to diagnose microphysics such like Kelvin-Helmholtz instabilities. 
Vikhlinin et al.\ (2001) observed a very sharp edge at the cold front in Abell~3667 and concluded that both diffusion and KHI were suppressed, invoking an amplified magnetic field. Recently, deep {\sl Chandra} exposures revealed a non-zero width at the cold front in the Virgo Cluster (Werner et al.\ 2016) implying the presence of KHI. 
Sanders et al.\ (2016) investigated the central region of the Centaurus Cluster. They observed a series of linear quasi-periodic structures which may be KHI rolls, although an interpretation involving AGN driven sound waves is also consistent with the data. 
As we will show, our deep {\sl Chandra} study of NGC~1404 is able to separate the observational evidence of diffusion and KHI for the first time.

\begin{figure*}[h]
   \centering
   \includegraphics[width=0.495\textwidth]{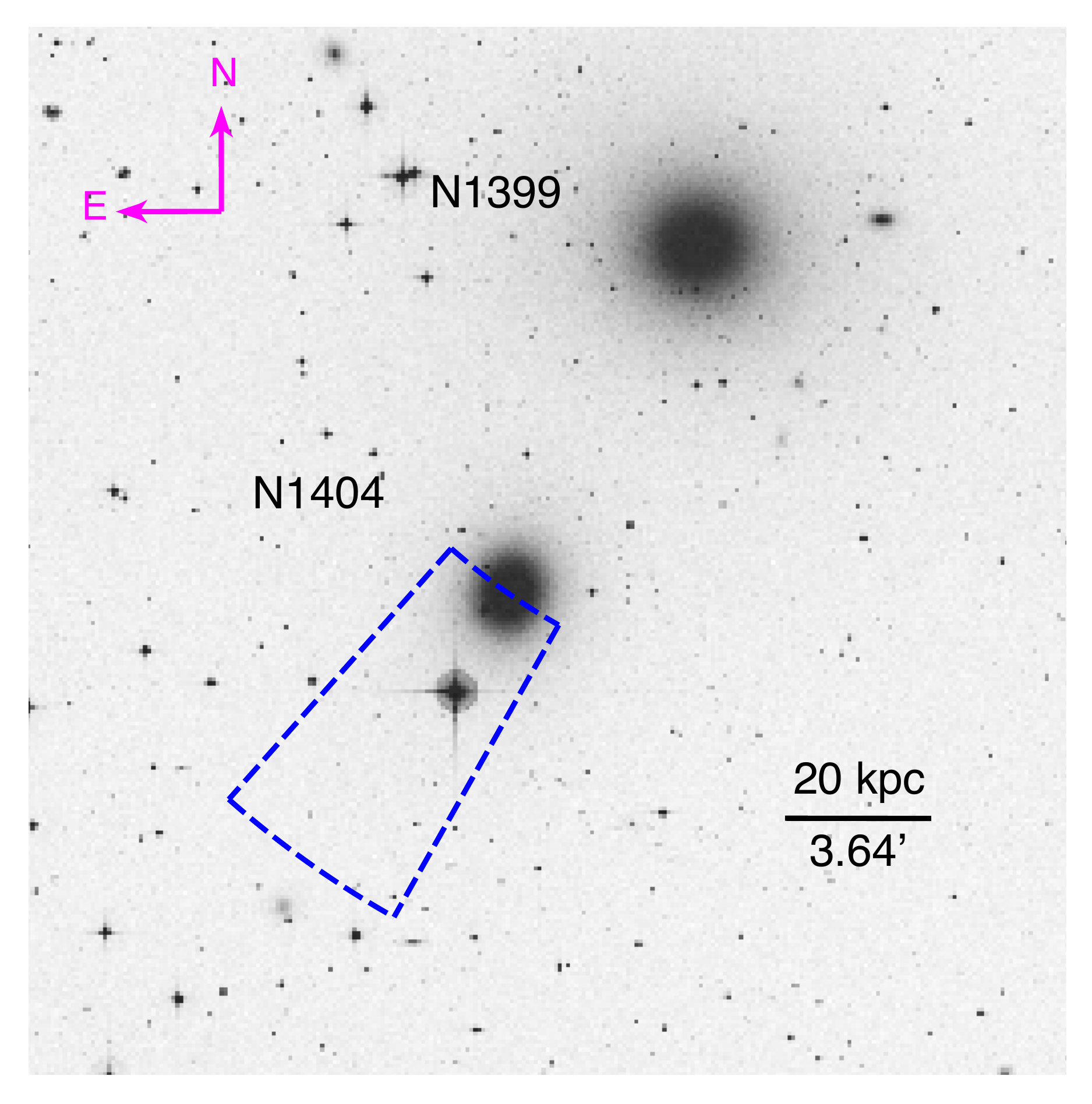}
    \includegraphics[width=0.495\textwidth]{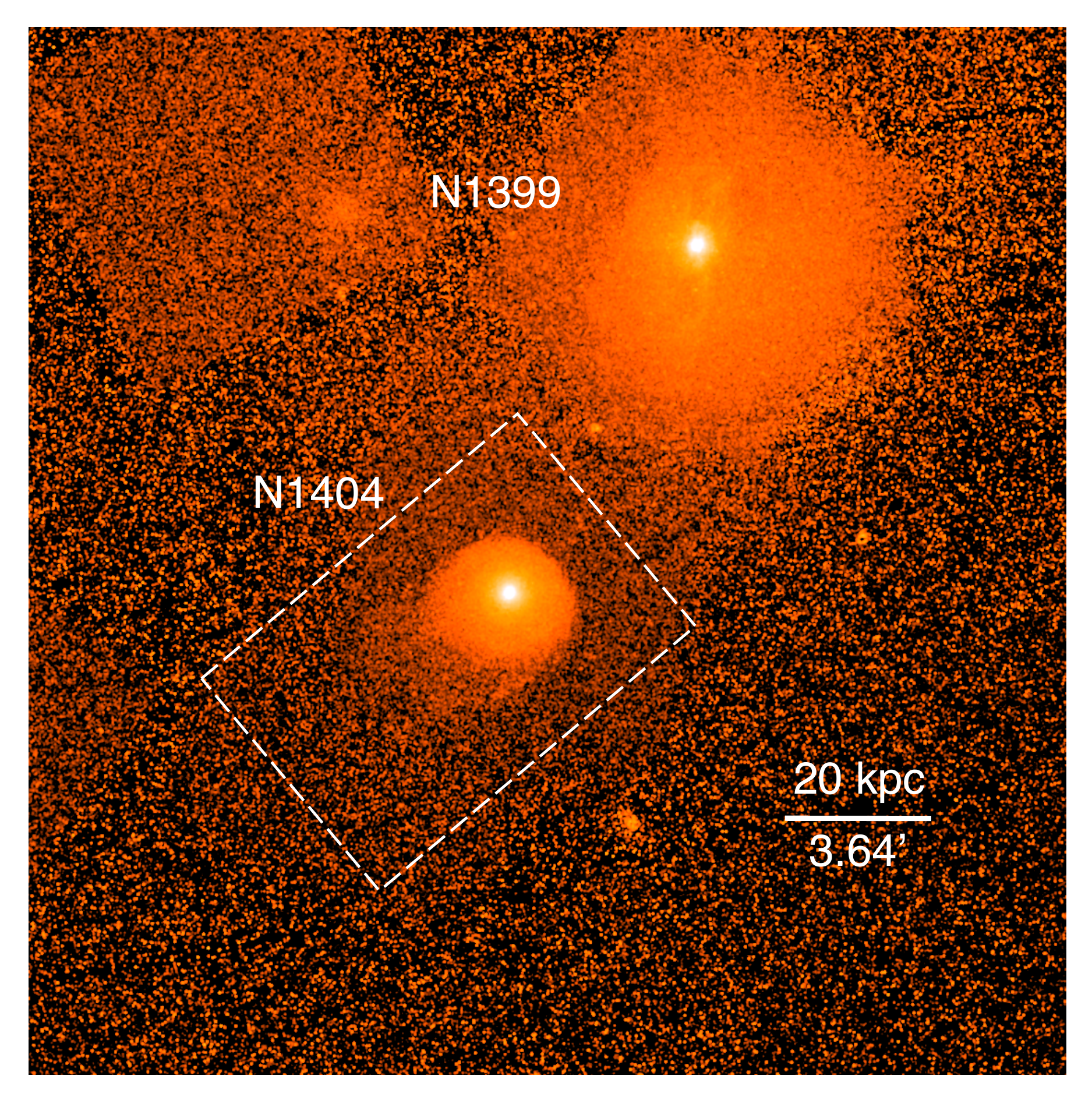} 
 \figcaption{\label{fig:N1399} {\it left:} Digitized Sky Survey (DSS) blue-band image of NGC~1404 and NGC~1399. Blue region marks the region used to derive the X-ray surface brightness profile of the stripped tail in Figure~\ref{fig:tail}. {\it right:} Exposure-corrected and blank-sky background subtracted {\sl Chandra} image of NGC~1404 and NGC~1399 in 0.5-2.0 keV. Box region marks the frame of Figure~\ref{fig:N1404}.}
\end{figure*}

NGC~1404 is a unique target to probe ICM microphysics. 
It is a bright elliptical galaxy falling through the ICM of the Fornax Cluster (centered on early-type galaxy NGC~1399) (see Figure~\ref{fig:N1399}). 
At a distance of $<$\,20 Mpc, contact discontinuities and small structures in this X-ray bright elliptical can be observed on scales below electron mean free path with {\sl Chandra}. The properties of the surrounding cluster gas, the medium in which any potential gas mixing or KHI would occur, can be well determined. 
In contrast to other well-known nearby bright infalling early-type galaxies (NGC4472--Kraft et al.\ 2011, M86--Randall et al.\ 2008, M89--Machacek et al.\ 2004; Roediger et al.\ 2015a,b),
NGC~1404 did not suffer interactions with other galaxies recently, nor is it disturbed by strong AGN outbursts.  
Its galactic atmosphere is the {\it cleanest} testbed to constrain transport processes, in that its data interpretation is not complicated by additional dynamical processes.

Previous studies of NGC~1404 revealed a sharp leading edge in the direction of NGC~1399 and an extended gaseous tail trailing behind (e.g., Machacek et al.\ 2005; Jones et al.\ 1997). 
In our first paper of a series of a very deep {\sl Chandra} observation of NGC~1404 (Su et al.\ 2016, hereafter Paper 1),
we determined that NGC~1404 resides in the same plane of the sky as the cluster center and is infalling at an inclination angle of $33^{\circ}$ with a Mach number of 1.32. The ISM of NGC~1404 and the ambient ICM have average thermal temperatures of 0.6 keV and 1.5 keV respectively.  
In this second paper, we focus on transport phenomena occurring between the ISM and ICM. 
This paper is organized as follows. The observations and data reduction are described in \S2. X-ray properties of the leading edge and the stripped tail are reported in \S3. In \S4 we discuss observational constraints on diffusion, conduction, viscosity and magnetic field. In particular, we present sub-kpc KHI rolls observed at the leading edge in \S4.3 and we inspect possible observational effect of magnetic draping in \S4.5. Our main conclusions are summarized in \S5.
Assumed cosmologies and the luminosity distance ($D_L=19$\,Mpc, $1^{\prime} = 5.49$ kpc, $z=0.00475$) are the same as Paper 1. 
Uncertainties reported in this paper are at 1$\sigma$ unless stated otherwise.

\section{\bf observations and data reduction}

We analyzed 670 ksec {\sl Chandra} observations of NGC~1404. We refer readers to Paper 1 for details of the observations and data preparation. 
In brief, all the data were reduced using {\sl CIAO}~4.8 tools following standard procedures. We produced blank-sky background subtracted, point source removed, and exposure corrected images for NGC~1404 and the Fornax Cluster. Readout artifacts were subtracted in both imaging and spectral analyses.
Blank-sky background was used for the spectral analyses of the ICM and local background was used for the ISM. 
In particular, we determined the spectral properties of a region just inside the contact edge, referred as Region ISM (T\,$=0.6$\,keV, $n_e=0.0061$\,cm$^{-3}$, and Fe\,$=0.52$\,Z$_{\odot}$), and of a region in the ICM offset from NGC~1404 but at the same distance relative to NGC~1399, referred to as Region ICM (T\,$=1.57$\,keV, $n_e=0.0012$\,cm$^{-3}$, and Fe$\,=0.30$\,Z$_{\odot}$). We fit the spectra with the model ${\tt phabs}\times{\tt vapec}$
in {\sl XSPEC} 12.7 using the C-statistic and the solar abundance standard of Asplund at al.\ (2006). Below, we present additional analyses to probe the ICM microphysics.

\subsection{\bf Imaging analyses}

\begin{figure}
   \centering
 \includegraphics[width=0.5\textwidth]{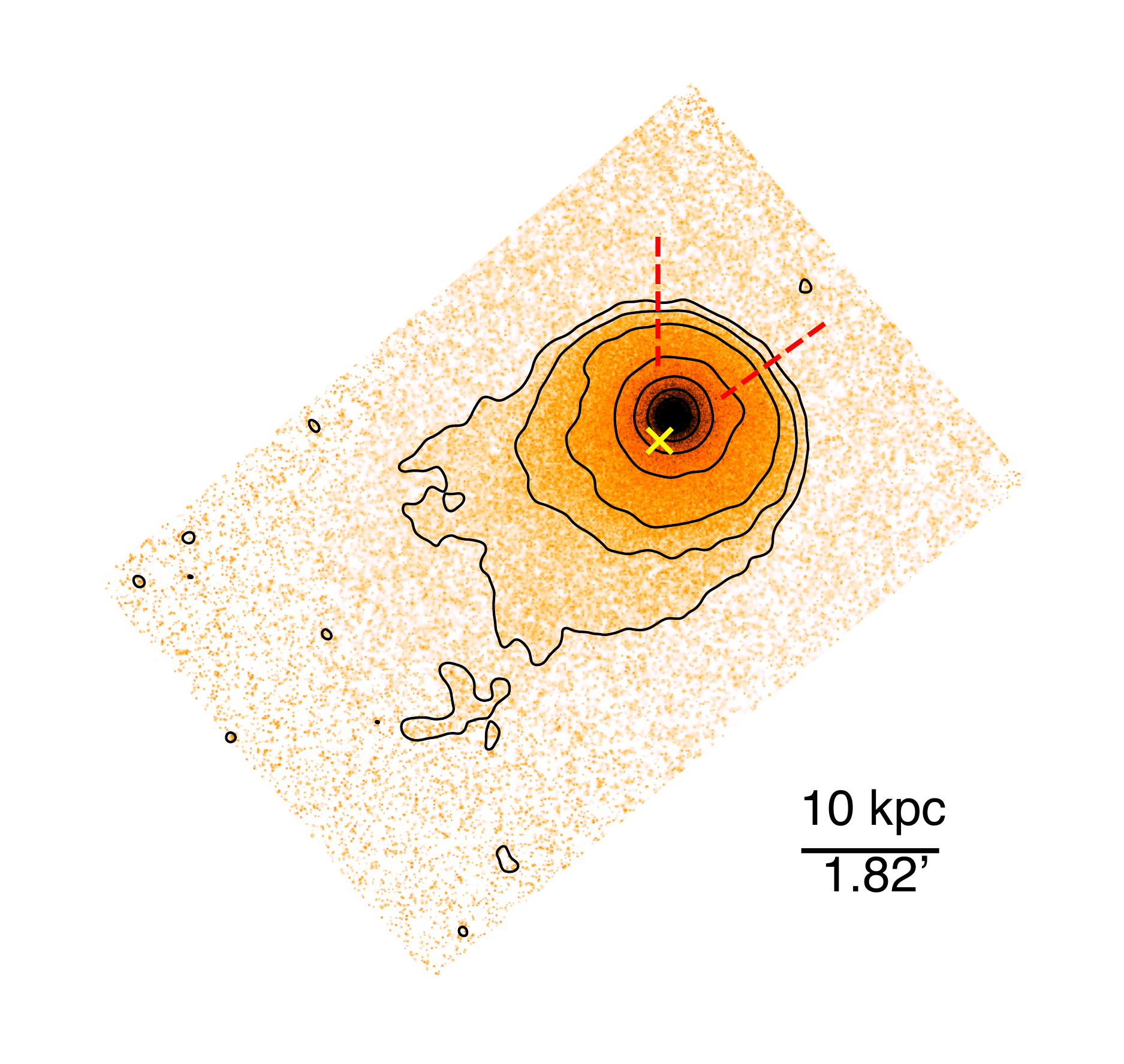}
\figcaption{\label{fig:N1404} {\sl Chandra} image of NGC~1404 in 0.7--1.3 keV. The image was exposure-corrected, blank-sky background subtracted, and smoothed using the CIAO tool {\tt dmimgadapt}. We identify the red sector as the sharpest edge (35$^{\circ}$--90$^{\circ}$), for which we derived the surface brightness profile. The ``x" sign (03h38m52.8s, -35d35m59.6s) is the center of the cold front curvature.}
\end{figure}

In Figure~\ref{fig:N1404},
we show an image of NGC~1404 restricted to the energy band of 0.7--1.3 keV which maximizes the ISM emission over the ambient cluster emission. 
To trace the X-ray emission contributed by the stellar diffuse emission, we derived the $K$-band surface brightness profile of NGC~1404 using the {\sl Two Micron All Sky Survey} ({\sl 2MASS}) (Skrutskie et al.\ 2006) archived image. The $K$-band accurately represents the old stellar populations of early-type galaxies. We made a 0.7--1.3 keV image, corresponding to the emission expected from unresolved X-ray sources in the old stellar populations seen in the $K$-band image of NGC~1404, which amounts to about 1\% of the total X-ray luminosity. Detailed analyses of {\sl 2MASS} data and conversion from the $K$-band data to X-ray were presented in Su \& Irwin (2013) and Su et al.\ (2015). 
Unresolved low mass X-ray binaries (LMXBs) also contribute to the diffuse X-ray emission. Its X-ray luminosity is found to correlate with stellar luminosity ($L_{\rm K}$) and globular cluster specific frequency ($S_{\rm N}$). 
Humphrey (2009) measured a low $S_{\rm N}$ of 1.8 for NGC~1404. Following the empirical relation (Boroson et al.\ 2011)
$L_{X}(\rm LMXB)=10^{28.88}\times{S_{\rm N}}^{0.334}$ ($L_{\rm K}/L_{\rm K,\odot}$) erg\,s$^{-1}$ for the 0.3--8.0 keV energy band, we obtained the corresponding 0.7-1.3 keV surface brightness by assuming a power law model with an index of 1.6 for LMXB (Irwin et al.\ 2003). This component is about three times the diffuse stellar emission.  
We also derived an 0.7--1.3 keV X-ray surface brightness profile of the Fornax cluster gas in the direction of NGC~1404. We fit this profile with a $\beta$ model profile 
We then subtracted the diffuse stellar emission, unresolved LMXB emission, and the ICM component from the 0.7--1.3 keV image of NGC~1404 to obtain the image best representing the gaseous ISM distribution of NGC~1404 shown in Figure~\ref{fig:N14042} (top).

\subsection{\bf Spectroscopic maps}

We performed a two-dimensional temperature analysis using Weighted Voronoi Tesselation (WVT) binning (Diehl \& Statler 2006),
a generalization of the Voroni binning algorithm presented in Cappellari \& Copin (2003). We first generated a WVT binning image containing 328 regions for the NGC~1404 image in the 0.7-1.3 keV band with a S$/$N of at least 30 in each bin. 
We used a single thermal {\tt vapec} component to model the hot gas emission in each region. We also included a power law component with an index of 1.6 to model the unresolved emission from LMXB (Irwin et al.\ 2003). The spectra for each region were fit to the model {\tt phabs}$\times$({\tt vapec}+{\tt pow$_{1.6}$}) using a blank-sky background. 
The abundance of each element was fixed to the best-fit of Region ICM (see Paper 1). 
The resulting temperature map is shown in Figure~\ref{fig:map} (left). 

\begin{figure}
   \centering
    \includegraphics[width=0.475\textwidth]{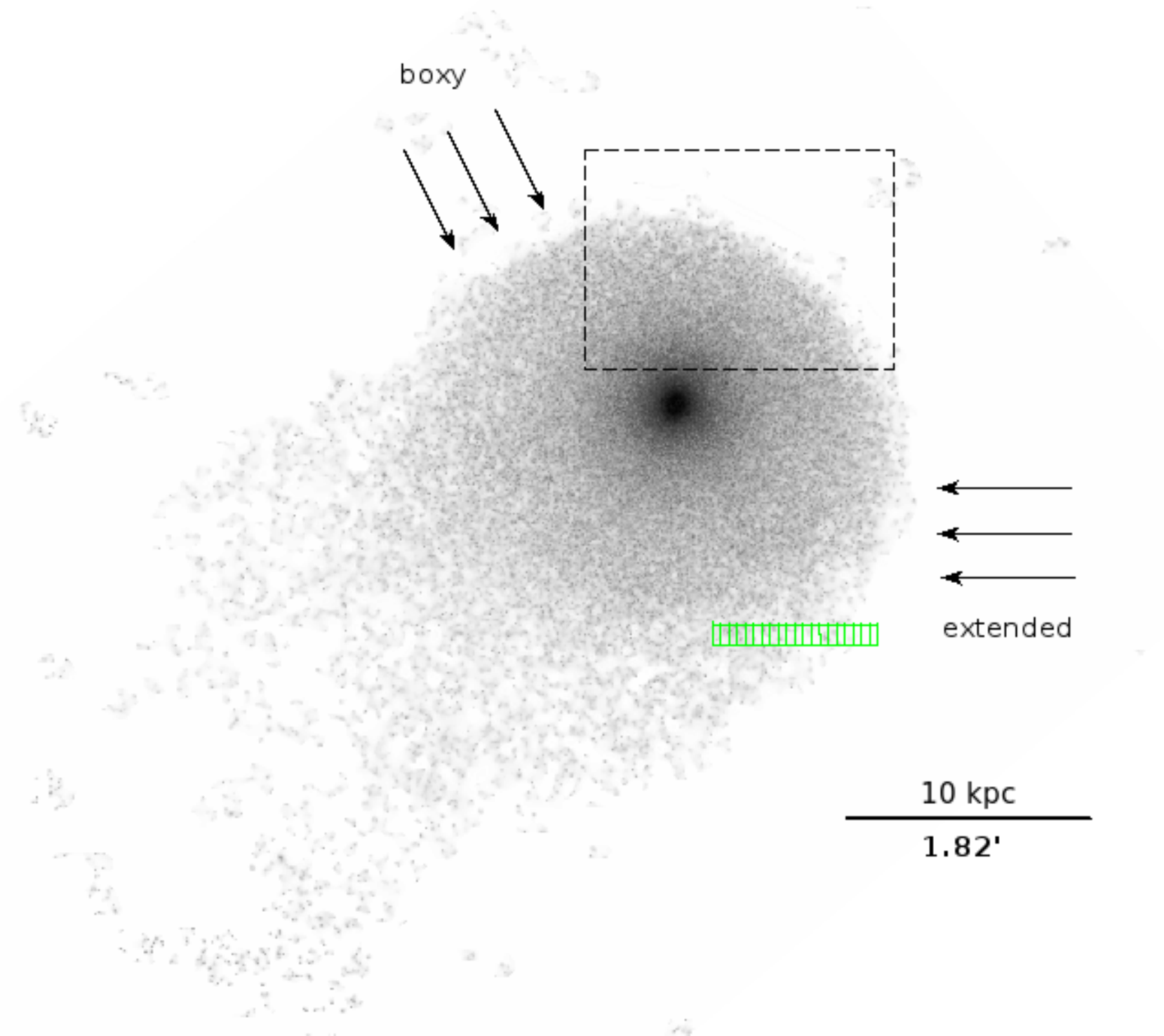} 
          \includegraphics[width=0.45\textwidth, frame]{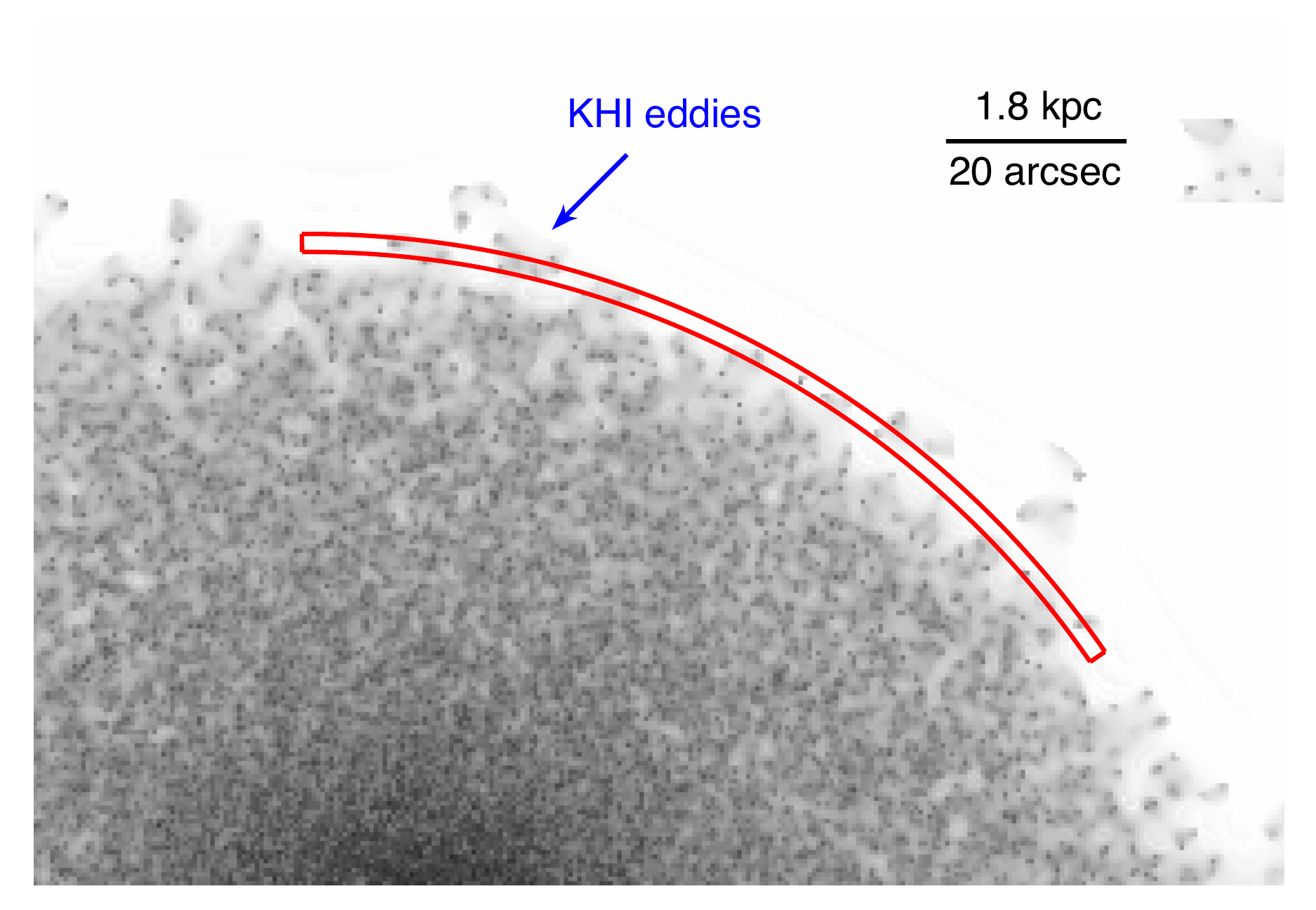}
      \includegraphics[width=0.45\textwidth]{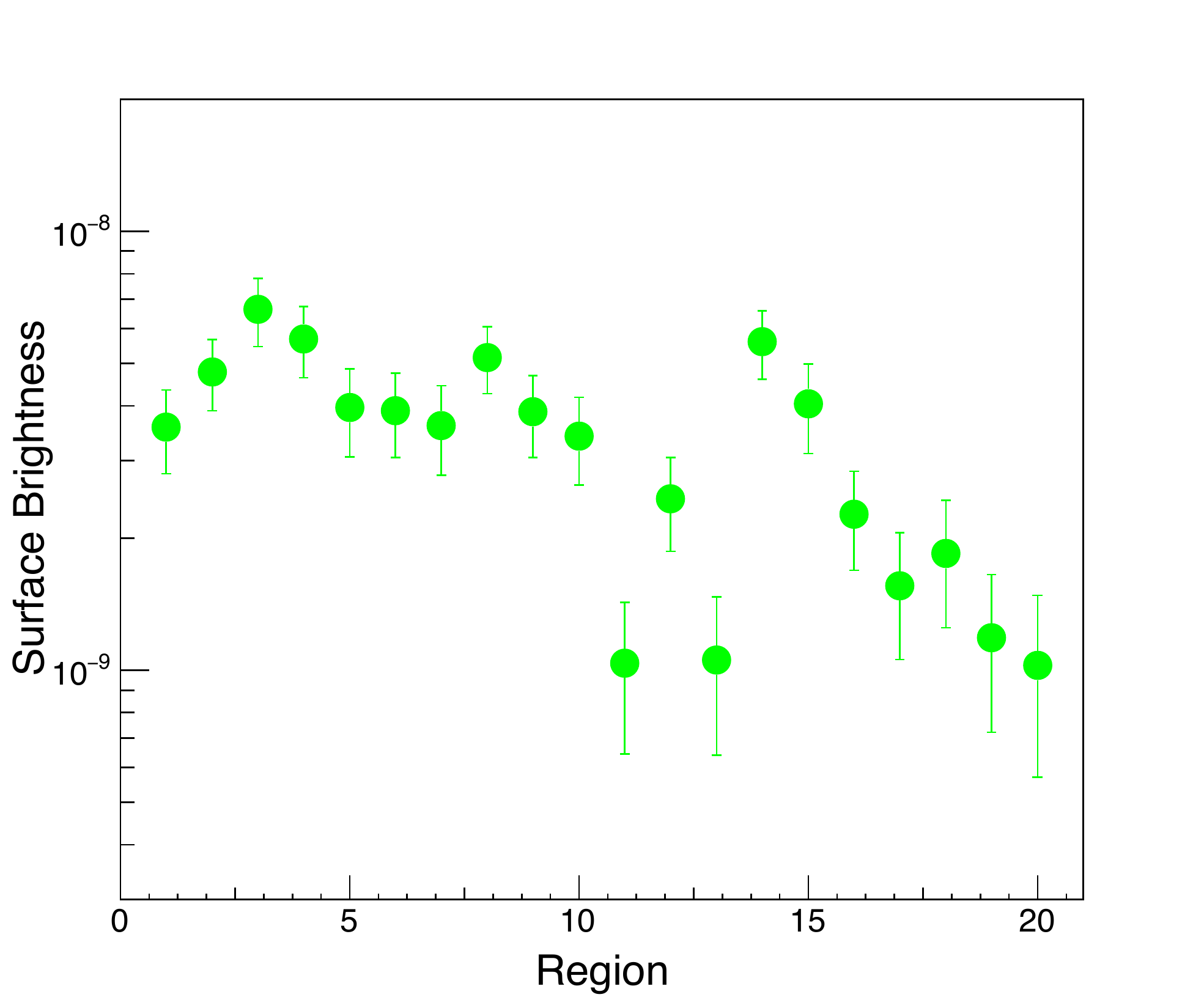}

\figcaption{\label{fig:N14042} 
{\it top:} Same as the Figure~\ref{fig:N1404} image but with the ICM, unresolved LMXBs, and diffuse stellar components subtracted. {\it middle:} a zoom-in image of the black box region in the {\it top} image. Red region ($2^{\prime\prime}$ wide) indicates the location of those ``eddies" found in the surface brightness profile in Figure~\ref{fig:pro}. 
{\it bottom:} surface brightness in downstream regions (green boxes in the {\it top} image; Regions 1--20: left to right).}
\end{figure}

We generated another WVT binning image containing 188 regions for the ISM of NGC~1404 using the ISM-only image (Figure~\ref{fig:N14042} top). S$/$N was also required to be $>$30 in each bin. The spectra of each region were fit to the same model, but local background was applied. 
The abundance of each element was fixed to the best-fit of Region ISM (see Paper 1). 
The resulting temperature map is shown in Figure~\ref{fig:map} (right).

\section{\bf Results}

\subsection{\bf Surface brightness and contact discontinuity}

\begin{figure*}[h]
   \centering
       \includegraphics[width=0.85\textwidth]{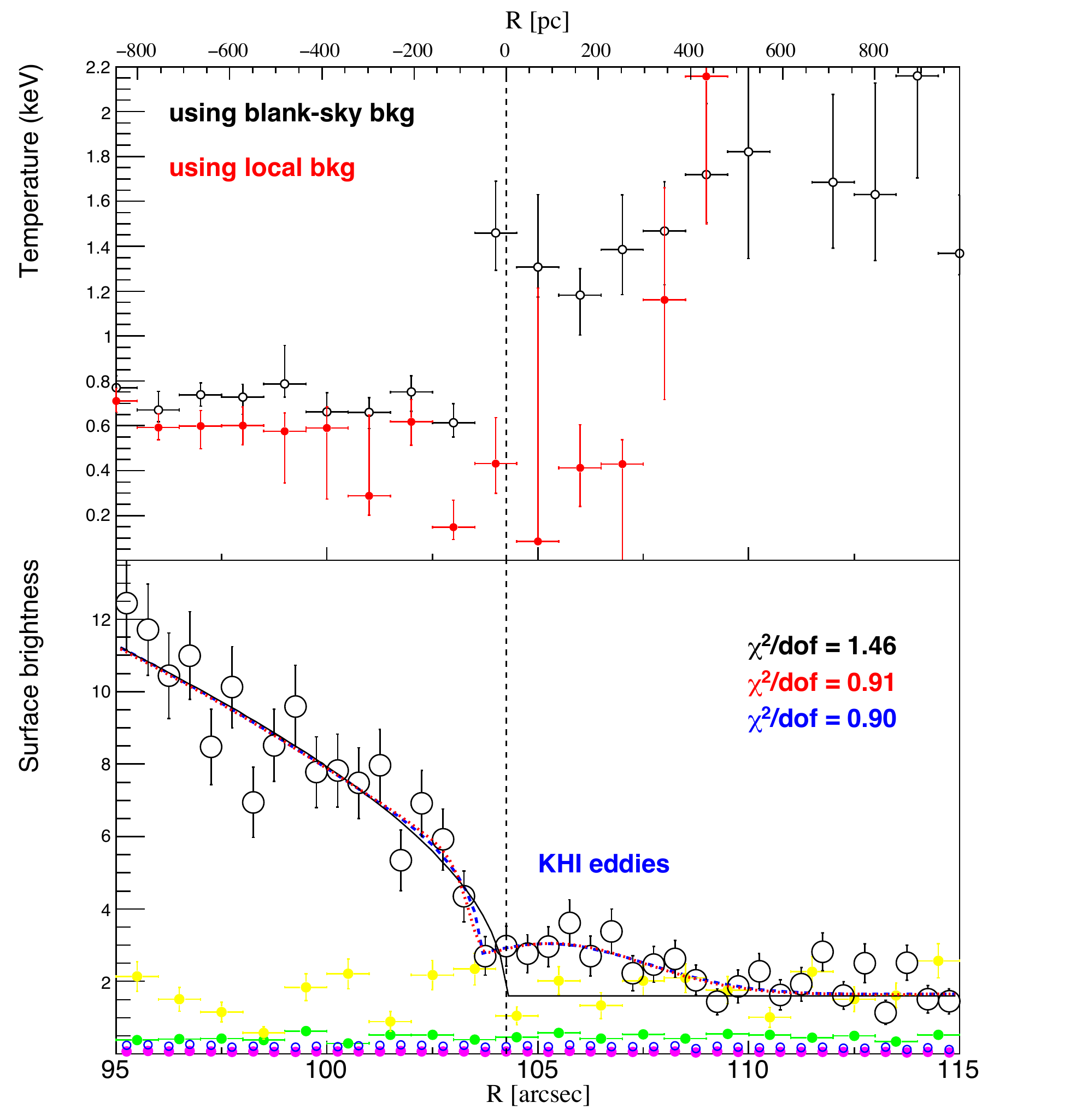}
\figcaption{\label{fig:pro} {\it top:} Temperature profile across the leading edge (35$^{\circ}$--90$^{\circ}$) of NGC~1404. Black profile was derived with blank-sky background and red profile was derived with local background. {\it bottom:} surface brightness profile in the 0.7--1.3 keV energy band (in units of $\times10^{-9}$ photons\,s$^{-1}$\,cm$^{-2}$\,arcsec$^{-2}$) across the leading edge (35$^{\circ}$--90$^{\circ}$) of NGC~1404. Yellow and green are surface brightness profiles in two other directions which are offset from NGC~1404 but at the same distance to NGC~1399. Black dashed line marks the location of the leading edge. Black solid line indicates the best-fit to our model containing both the Fornax ICM component and the NGC~1404 ISM component. The ``eddies" exceeding the background can be identified just outside the leading edge; blue dashed line indicates a model including an additional gaussian component for this eddy feature; red dashed line indicates a further modified model with the edge convolved with a gaussian component. 
Blue and magenta circles: expected surface brightness profile contributed by unresolved LMXB and stellar emission respectively.}
\end{figure*}

As shown in Figure~\ref{fig:N1404}, we identify the leading edge of NGC~1404 as its sharpest edge (35$^{\circ}$--90$^{\circ}$), for which we obtained a surface brightness profile using a radial bin width of $0\farcs 5$ (45\,pc), plotted in Figure~\ref{fig:pro}.
This surface brightness profile was fit to a model containing both the Fornax ICM component ($\beta$ profile) and the NGC~1404 ISM component (power-law density profile) as described in Paper 1. The best-fit model, indicated as a black solid line in Figure~\ref{fig:pro}, gives a density jump of $5.2\pm0.2$ at $r=104\farcs25\pm0.02$ (centered on the yellow cross in Figure~\ref{fig:N1404}). 
The contact discontinuity at this boundary does not show any broadening larger than the instrumental spatial resolution. 
We note some eddy feature in the surface brightness profile, in excess of the best-fit model, just outside the edge. The F-test probability for the significance of this eddy feature (a gaussian component) is $1.96612\times10^{-7}$ ($5.2\,\sigma$). The fit is not further improved when we convolve the edge with an additional gaussian component. 
The best-fit gaussian broadening has a standard deviation of $0\farcs 37$ (33 pc), making the front width $w \simeq \sqrt{2\pi} \sigma\simeq2.5 \sigma = 83$ pc (Churazov \& Inogamov 2004).
We compared the surface brightness in the eddy feature to the emission from the Fornax cluster (to the east and west of NGC~1404) at the same distance from its center,
as indicated in yellow and green in Figure~\ref{fig:pro}. The surface brightness of the ``eddies" is in excess of the ICM background. A detailed view of this feature on the {\sl Chandra} image is shown in Figure~\ref{fig:N14042} (middle); note that this feature is not on any CCD edge/gap of any of these observations.

\begin{figure*}[h]
   \centering
       \includegraphics[width=1.0\textwidth]{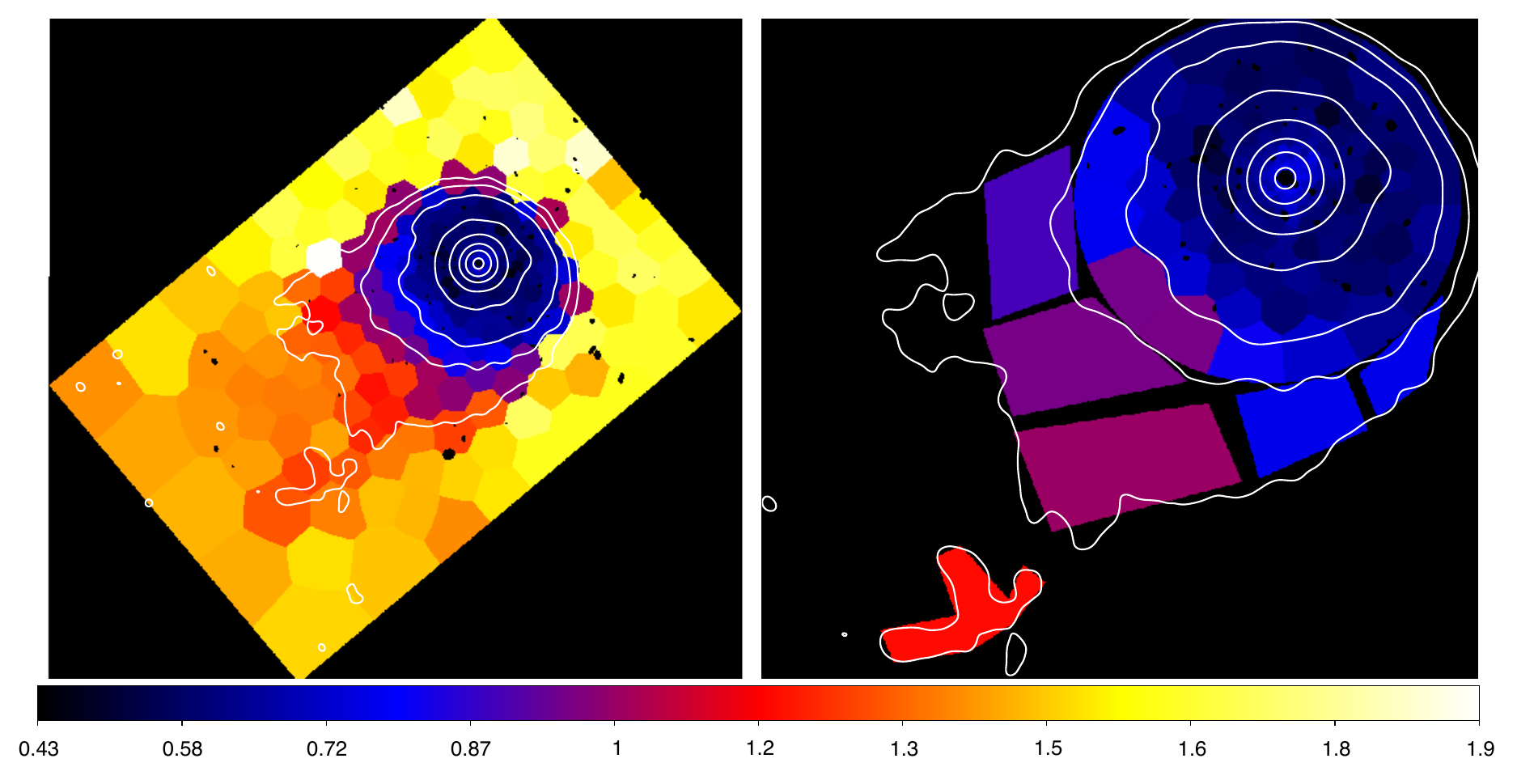}
          \vspace{1.cm}
\figcaption{\label{fig:map} {\it left:} temperature map of NGC~1404 and its ambient ICM derived with blank-sky background. {\it right:} temperature map of NGC~1404's gaseous ISM (remnant core plus stripped tail) derived with local background. Both maps are in units of keV. White contours represent {\sl Chandra} X-ray distribution of NGC~1404's gaseous ISM.}
\end{figure*}

To study the nature of this eddy feature, we derived the temperature profiles across the edge using blank-sky background (black) and local background (red) separately as shown in Figure~\ref{fig:pro}.
The width of each annulus bin is 1$^{\prime\prime}$ (90\,pc). We used a single thermal {\tt vapec} model for the spectral fit since these regions are beyond the optical extent of NGC~1404 and the contribution from unresolved LMXBs should be negligible. Metal abundances were fixed 
to those of Region ICM with blank-sky background, and of Region ISM with local background.
Temperature profiles of regions of enhanced surface brightness (such as the NGC~1404's atmosphere and the ``eddies") can be best described when the local background is employed, since the contamination from the cluster hot gas should be eliminated. 
The eddy feature shows a temperature as low as the ISM temperature with a gradient increasing outward and eventually reaching the ICM temperature.  
The temperature derived with the blank-sky background reflects the temperature of the Fornax ICM outside the edge, and some average of the ICM and ISM projected inside the edge. 
The temperature varies from $\sim0.6$\,keV to $\sim1.5$\,keV within 200 pc across the cold front, implying an ineffective conduction at the leading edge.

\subsection{\bf The stripped tail}

\begin{figure}[h]
   \centering
       \includegraphics[width=0.5\textwidth]{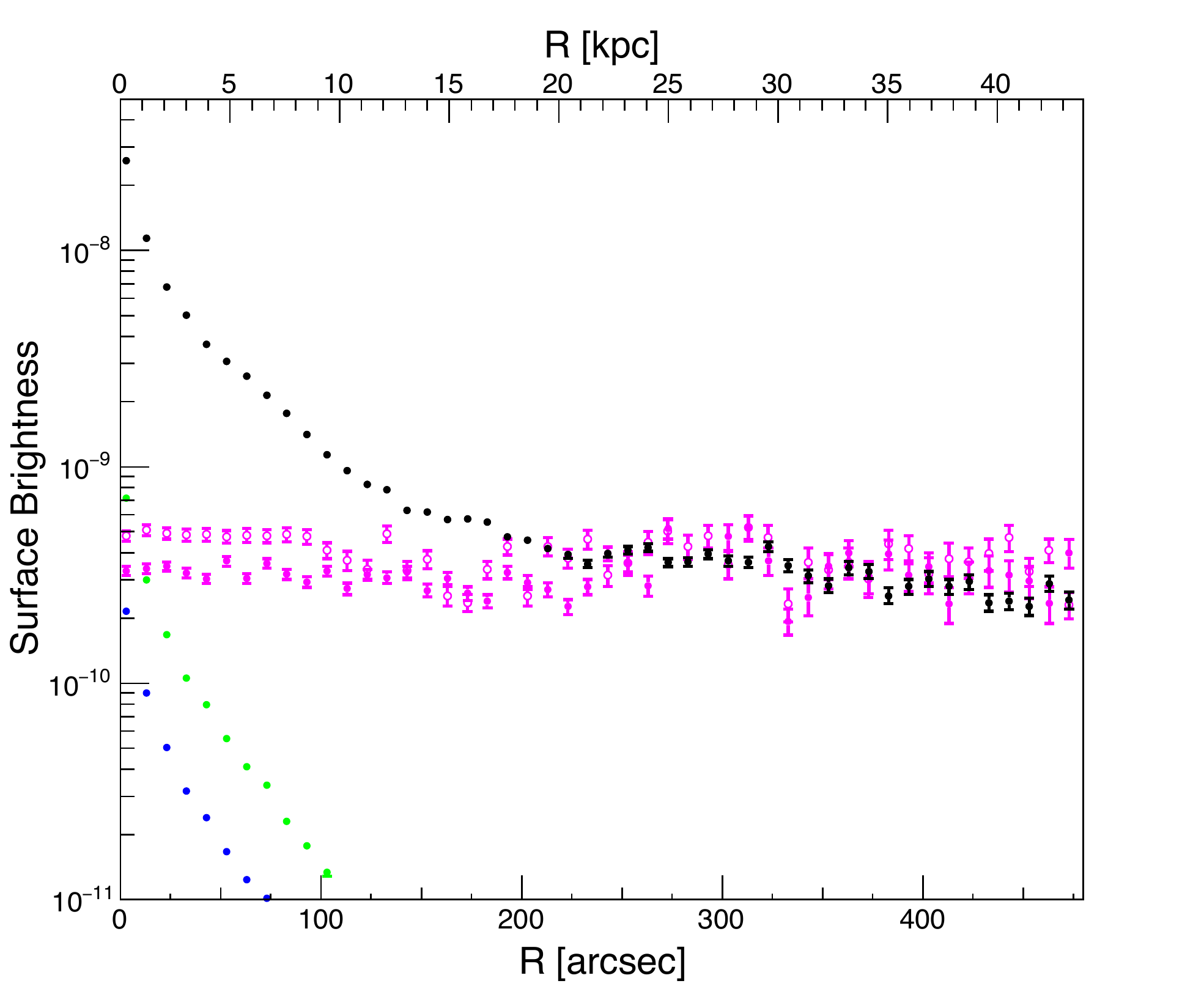}
\figcaption{\label{fig:tail} Black dots: surface brightness profile (in units of photons\,s$^{-2}$\,cm$^{-2}$\,arcsec$^{-2}$) in the energy band of 0.7-1.3 keV extracted in rectangular sections from the galaxy center to downstream regions (blue box in Figure~\ref{fig:N1399}-left). Open and solid magenta circles: surface brightness profiles sampled in two other directions which are offset from NGC~1404 but at the same distance to NGC~1399. Green and blue dots: expected surface brightness profiles contributed by unresolved LMXB and stellar emission respectively.}
\end{figure}

The extended stripped tail of NGC~1404, trailing the core, is evident in the contours on the {\sl Chandra} X-ray image and can be clearly seen in the image with the ICM component subtracted (Figure~\ref{fig:N14042} top). In Figure~\ref{fig:tail}, we compared the surface brightness profile of the galaxy downstream to that of the ambient ICM. This tail is as wide as the remnant core (16 kpc) and its length is 8 kpc in projection, appearing short even compared with X-ray stripped tails observed in highly line-of-sight infalling early-type galaxies (e.g., M86--Randall et al.\ 2008; NGC~1400--Su et al.\ 2014).

We extracted spectra from a few regions in the stripped tail of NGC~1404 and fit them to a single thermal {\tt vapec} model; a local background was used. Abundances were fixed at the best-fit values for Region ISM (Fe\,$=0.52$\,Z$_{\odot}$, see \S2.1). 
We compared the temperatures of these regions to that of the remnant core in Figure~\ref{fig:map} (right). 
The tail gas, at a temperature of 0.9--1.0 keV, is consistently hotter than the gas in the remnant core ($<0.6$ keV) and cooler than the ambient ICM ($\sim1.5$ keV). This is suggestive of thermal conduction heating the stripped gas or turbulent mixing of hotter ICM occurring in the downstream tail.  

One would expect the stripped tail to have a metallicity similar to the remnant core. 
We fit the spectra extracted from the entire tail, leaving metallicities free to vary. We obtain best fits of k$T=0.91^{+0.03}_{-0.03}$\,keV and Fe = $0.29^{+0.21}_{-0.09}$\,$Z_{\odot}$. Its metallicity is as low as that of Region ICM of Fe\,$=0.30$\,Z$_{\odot}$. This may be due to ``Fe-bias," caused by fitting a single thermal model to multiple phase gas (hot ICM and cool ISM) as described by Buote (2000). We therefore applied a model containing two temperature components but 
find the data insufficient to constrain the metallicities. 
We further set one component to have a temperature twice that of the other based on the average ratio of the two components found in the ISM of early-type galaxies (Su \& Irwin 2013; Su et al.\ 2015). The resultant best-fit Fe rises to $0.47^{+0.61}_{-0.17}$\,$Z_{\odot}$ but is not significantly different from the value found in the single temperature fit.

\section {\bf Discussion}

We resolve substructures in surface brightness and temperature of NGC~1404 down to $<$\,100\,pc, 
thanks to the combination of the proximity of NGC 1404, the superb spatial resolution of {\sl Chandra}, a very deep (670 ksec) exposure, and the low temperature of NGC~1404. Based on observational facts, we describe the transport processes and probe the microscopic properties of the cluster plasma.

\subsection{\bf Diffusion and conduction}

The surface brightness profile of the leading edge reveals a sharp density discontinuity at the front. Here, we compare the width of the discontinuity to the characteristic mean-free-path (mfp) of electrons, $\lambda_e$, in the hot plasma on both sides of the leading edge. Sarazin (1988) gives:
\begin{equation}
\lambda_e=\frac{3^{3/2} (kT_e)^2}{4\pi^{1/2}n_e^4\rm ln \Lambda},
\end{equation}
where the Coulomb logarithm is \begin{equation}\rm ln \Lambda = 37.8+ ln \left[\left(\frac{T_e}{10^8 K}\right)\left(\frac{n_e}{10^{-3} cm^{-3}}\right)^{-1/2}\right].\end{equation}
For densities and temperatures in Region ISM and Region ICM,
the mfp is $\lambda_{\rm in} = 20$\,pc in the ISM and $\lambda_{\rm out} = 653$\,pc in the ICM. The effective mfp across the front can be estimated as (Vikhlinin et al.\ 2002)

\begin{equation}
\lambda_{\rm in\rightarrow out} = \lambda_{\rm out} \frac{T_{\rm in}}{T_{\rm out}}\frac{G(1)}{G(\sqrt{T_{\rm in}/T_{\rm out}})},
\end{equation}
and
\begin{equation}
\lambda_{\rm out\rightarrow in} = \lambda_{\rm in} \frac{T_{\rm out}}{T_{\rm in}}\frac{G(1)}{G(\sqrt{T_{\rm out}/T_{\rm in}})},
\end{equation}
where $G(x)=[\Phi(x)-x\Phi^{\prime} (x)]/2x^2$ and $\Phi(x)$ is the error function. In the case of NGC~1404 and its ambient ICM, we obtain $\lambda_{\rm in\rightarrow out} = 287$\,pc and $\lambda_{\rm out\rightarrow in} = 70$\,pc.

The surface brightness profile of the front is consistent with no broadening at the resolution limit of the instrument (Figure~\ref{fig:pro}). Therefore, the width of
the edge cannot be much larger than the resolution of $0\farcs5\simeq45$\,pc, that is $<100$\,pc, which is also consistent with the gaussian smoothing result.
The width of surface brightness discontinuity is smaller than the mean free path across the front ($\lambda_{\rm in\rightarrow out}$), which implies a suppressed electron diffusion.

Heat diffusion is the fastest transport process, as it is dominated by the rapidly moving electrons of a plasma in equipartition. Heat flows from the ICM to the ISM (out $\rightarrow$ in).  
In the ICM, it
will reduce the temperature and raise the density while in the ISM it has the opposite
effect, as required to maintain a roughly constant pressure across the front.  Because the temperature scale height is smaller than the electron mean free path at the front, the heat flux would saturate at the value (Cowie \& McKee 1977)
$q_{\rm sat}=5\Phi_{\rm s} \sqrt{P/\rho}P=\frac{5\Phi_{\rm s}c_s}{\sqrt{\gamma}}P$, where $\Phi_{\rm s}=1.1$ for a fully ionized gas.  
This translates to an effective speed of the saturated heat flow of $\simeq 4.3\,c_s$. Since this speed is substantially greater than the flow speed, we should expect to see the effects of isotropic heat conduction at the front, at least on the scales of the electron mean free path in the ISM and the ICM. 
The radial bin size we used for the temperature analysis is 1$^{\prime\prime}\simeq90$\,pc, smaller than $\lambda_{\rm out}$. Temperature profiles display abrupt changes; strong gradients on a scale of $\lambda_{\rm out}$ ($>500$\,pc) are not observed in the ICM (Figure~\ref{fig:pro}-top).
This result implies a suppressed thermal conduction in the ICM. Admittedly, we would need simulations to get a more accurate constraint on the conduction.

\subsection{\bf Kelvin-Helmholtz instabilities}

We observe an eddy feature of enhanced surface brightness just outside the leading edge at a significance of 5$\sigma$. Figure~\ref{fig:pro} (top, red) indicates this feature has a low temperature suggesting that the brightness enhancement must be caused by gas originating from within the ISM rather than an ICM surface brightness fluctuation. 
In Paper 1 we estimated that the apparent leading edge in projection is offset from the actual stagnation point by 33$^{\circ}$. 
The most plausible interpretation of this enhancement is that KHI have developed on the side of this galaxy. 
We show the location of this eddy feature on the {\sl Chandra} image of NGC~1404 in Figure~\ref{fig:N14042} (middle). The red annulus sector is in the same direction ($35^{\circ}$--$90^{\circ}$) as extraction regions used in Figure~\ref{fig:pro}. Its width is 2$^{\prime\prime}$ (180\,pc) four times as wide as that in Figure~\ref{fig:pro}, covering the entire low temperature eddy feature. One or two sub-kpc patches of enhanced surface brightness in this layer stand out. We identify them as possible KHI rolls. 
We note that surface brightness profiles within the contact edge also shows wiggling. This is not surprising given that KHI ought to develop around almost the entire surface of the galaxy.

We identify potentially larger KHI rolls on both sides of the leading edge in Figure~\ref{fig:N14042} (top). 
To the east of the leading edge, there is a $\sim$\,5 kpc wide section of the front that is ``boxy," rather than following the smooth arc of the leading edge. Similar boxy sections can be seen in multiple sloshing cold fronts in Abell~496 (Dupke et al.\ 2007). The origin of such boxy morphology has been unknown until recently. Roediger et al.\ (2012) reproduced these features through high-resolution, non-viscous simulations and demonstrated that they arise from KHI combined with projection effects. 
The edge to the west of the leading edge appears less sharp and is also $\sim$\,5 kpc wide. Its surface brightness profile outside the best-fit edge exceeds the ambient ICM for 2 kpc, for which we measure a temperature similar to that of the ISM, consistent with the properties of KHI rolls.   
These features, to the east and west, are $\sim$70$^{\circ}$ away from the actual stagnation point where we do expect larger KHI rolls due to the larger shear flow velocity. 
In examining regions further downstream (indicated by green boxes in
Figure~\ref{fig:N14042} top) we note that an obvious deficit is present in
the surface brightness distribution (Figure~\ref{fig:N14042} bottom). 
This deficit is indicative of the initiation of the KHI roll-galaxy
bifurcation, eventually leading to the complete stripping of the KHI
rolls into the ICM.

\subsection{\bf Viscosity}

Viscosity can inhibit the growth of KHI by damping out perturbations less than a certain wavelength. 
The presence of KHI in this system allows us to put quantitative constraints on the viscosity of the ICM. 
The growth of KHI at any given shear layer would be suppressed if the Reynolds number, Re, is smaller than a critical value Re$_{\rm crit}$.
The Reynolds number is defined as ${\rm Re}=\frac{\ell U}{\nu}$, where $U$ is the shear flow velocity, $\ell$ is a characteristic length scale, and $\nu$ the viscosity. When the particle mean free paths are determined by Coulomb collisions, the viscosity scales as $\nu \propto T^{5/2}n^{-1}$ (Spitzer 1956).   
For ICM of temperature $T_{\rm ICM}$ and electron density $n_{\rm e,ICM}$, 
the Reynolds number becomes 
$$
{\rm Re}=10{f_{\nu}}^{-1}\left(\frac{U}{400\,\rm km/s}\right)\left(\frac{\ell}{10\, \rm kpc}\right)
$$
\begin{equation}
~~~~~~~~~~\times\left(\frac{n_{\rm e, ICM}}{10^{-3}\rm cm^{-1}}\right)\left(\frac{kT_{\rm ICM}}{\rm 2.4 \,keV}\right)^{-5/2}, 
\end{equation}
where $f_\nu$ is the fraction of the viscosity relative to the Spitzer value.
Following Roediger et al.\ (2013), we can calculate the critical Reynolds number
${\rm Re}_{\rm crit}=16\sqrt{\Delta}$, where  $\Delta = \frac{(\rho_{\rm hot}+\rho_{\rm cold})^2}{\rho_{\rm hot}\rho_{\rm cold}}$ characterizes the contrast of the gas densities on each side of the interface.

We observe the height of KHI as small as $h=180$\,pc. 
Numerical simulations show that the characteristic length scale can be a few times the height of KHI. Here, we take 
$\ell\simeq3h$ as a typical ratio as demonstrated in Roediger et al.\ (2013). This requires  ${\rm Re}>{\rm Re}_{\rm crit}$ for $\ell\approx0.5$\,kpc. 
We take $U=v^{\prime}=1.1 \sin\theta^{\prime} c^{\prime} = 446$\,km\,s$^{-1}$ (where $\theta^{\prime} =33^{\circ}$ and $c^{\prime}=\sqrt{\frac{T^{\prime}}{T_1}}c_s=1.15c_s$, see Paper 1).
From our analysis, we take $n_{\rm e, ICM}=0.0012$ cm$^{-3}$, $T_{\rm ICM}= 1.57$ keV, $\rho_{\rm cold}\propto 0.0061$ cm$^{-3}$, and $\rho_{\rm hot}\propto$ 0.0012 cm$^{-3}$. 
In this case, we find $f_\mu < 0.05$, so that the viscosity is no more than 5\% of the Spitzer value.

The length of the stripped tail can provide an independent diagnostic of the plasma viscosity.  
Through hydrodynamic simulations, Roediger et al.\ (2015a) predict that a stripped tail in a plasma of high viscosity should be as cold as the remnant core and extend out to 30 kpc behind the galaxy, while in the case of an inviscid plasma, the stripped tail is mixed efficiently with the ICM.  
In practice, however, the remnant core of a galaxy can temporarily shield a stripped tail. In other words, a long extended tail can be expected in newly infalling galaxies even in a fully turbulent plasma. Fortunately, NGC~1404 has entered the Fornax Cluster long ago and the shielding effect is by now greatly reduced (see Paper 1).    
Thus it is appropriate to use the tail length of NGC~1404 to diagnose the viscosity of the plasma. 
For an inclination angle of 33$^{\circ}$, the actual tail length is 10\,kpc, remarkably short compared with typical stripped tails of early-type galaxies (M86--Randall et al.\ 2008; NGC~1400--Su et al.\ 2014). 
This result favors a low viscosity plasma.

\subsection{\bf Instability limited by intrinsic width}

Churazov \& Inogamov (2004) argue that a finite intrinsic width of the interface may be adequate to stabilize the cold front. 
Consider an interface between two phases of fluids. The velocity grows along the interface $v=v_0 \sin\theta$. The wave number $k$ (wavelength $2\pi/k$) of the perturbation decreases correspondingly $k(\theta)=k_1\frac{v_1}{v}=k_1\frac{\sin\theta_1}{\sin\theta}$, where $\theta_1$ is the initial position.
The growth of the perturbation propagating along the interface can be obtained through the Wentzel-Kramer-Brillouin approximation 
\begin{equation}
\delta = \delta_1 \left(\frac{\sin\theta_1}{\sin\theta}\right)^2 {\rm exp} \Big\{\int \gamma(t){\rm d}t \Big\},
\end{equation}
where $\delta_1$ is the initial amplitude and $\gamma$ is the increment of KHI.
Any real interface, no matter how thin, has a finite width. The thickness of the interface, $w$, is set by the diffusion process. Only perturbations with wave number $k \leq k_{\rm max}\sim 1/w$ can become unstable. 
This sets a minimum angle
at which KHI starts to develop: $\theta_{\rm min}=\frac{k_2}{k_{\rm max}}$sin$\theta_2$, where $k_2$ is the wave number measured at $\theta_2$. 
We take the form given by Rayleigh (1945) $\gamma(k,w)=\frac{v}{2w}\sqrt{e^{-2kw}-(kw-1)^2}{\rm d}\theta$, where $k\leq1.2785/w$ is the condition for instability. 
The growth factor for a perturbation starting at $\theta_1$ and measured at $\theta_2$ is (Equation (11) in Churazov \& Inogamov 2004)

$$
{\rm growth ~factor}=\left(\frac{\sin\theta_1}{\sin\theta_2}\right)^2 
$$
\begin{equation}
{\rm exp}\Big\{\frac{R}{w}\frac{1}{\sqrt{\Delta}}\int_{\theta_1}^{\theta_2}\sqrt{e^{-2kw}-(kw-1)^2}{\rm d}\theta\Big\},
\end{equation}
where $\Delta$ is the density ratio of the two fluids and $\theta_1=\theta_{\rm min}$.

\begin{figure}[h]
   \centering
       \includegraphics[width=0.5\textwidth]{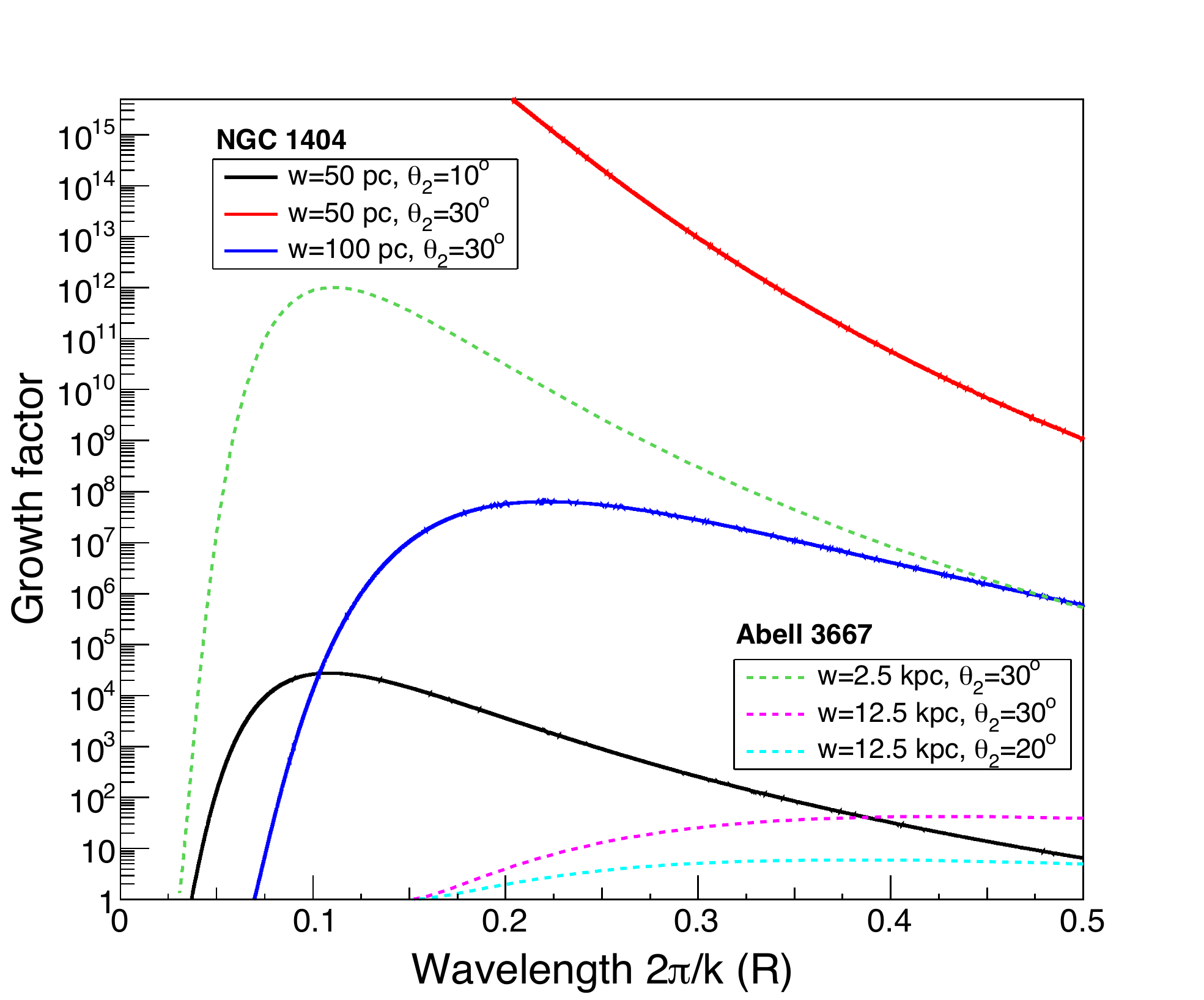}
\figcaption{\label{fig:gf} Growth factor as a function of wavelength in units of front curvature radius. Solid lines and dashed lines are for several parameter values for NGC~1404 and Abell~3667 respectively. Blue line and magenta lines represent their best-fit parameters.}
\end{figure}

This theory was applied to the best studied cold front in Abell~3667 (Churazov \& Inogamov 2004). Given the upper limit of the interface thickness relative to the curvature radius of the front, the growth factor of KHI is overall sufficiently small (magenta dashed line in Figure~\ref{fig:gf}) indicating that there is no need to invoke dynamically important magnetic fields in the case of Abell~3667.  
Churazov \& Inogamov (2004) stressed that the growth factor is critically dependent on 
the width of the interface. The upper limits of this width are $\sim10$\,kpc and $\sim$100\,pc for Abell~3667 and NGC~1404 respectively, mostly because Abell~3667 is $>10$ times more distant.  
Here, we calculate the growth factor as a function of wavelength for the cold front in NGC~1404 (blue solid line in Figure~\ref{fig:gf}). We take $R=8$\,kpc, $\theta_2=30^{\circ}$, $w=100$\,pc, and $\Delta=5$. Its growth factor stays above $10^{6}$ for nearly all wavelengths. 
The effect of intrinsic width is much more prominent in Abell~3667 than in NGC~1404.

\subsection{\bf The magnetic fields}

Magnetic field can effectively suppress diffusion. 
The effective mfp for electron diffusion perpendicular to the magnetic field is on the order of ${r_g}^2/\lambda_e$ (Spitzer 1956), where 
$
{r_g}=(\frac{T}{10^8 {\rm K}})^{1/2}(\frac{m}{m_e})^{1/2}(\frac{B}{1\mu\rm G})^{-1}\times10^{-10}\,{\rm pc}
$
is the gyroradius determined by the electron temperature and the magnetic field strength (Sarazin 1988). At the leading edge, the effective mfp should be below the spatial resolution of our analysis (45 pc), which sets a lower limit on the magnetic field strength of 10$^{-12}\mu$G. 

Magnetohydrodynamic simulations show that
magnetic field can be swept up at the leading edge of the ISM in a galaxy moving through the ICM, draping magnetic field over the interface between the ISM and the ICM (Dursi \& Pfrommer 2008); 
in effect, charged particles can only diffuse along the field lines.
This process can in principle suppress the growth of KHI. 
Simulated sloshing showed that 
KHI develops at sloshing fronts in galaxy clusters and found that its growth can be inhibited by a strong magnetic field (ZuHone et al.\ 2011). 
Vikhlinin et al.\ (2002) argue that the total magnetic pressure needs to meet the following condition to suppress KHI. 
\begin{equation}
P_{\rm mag}=\frac{B_c^2+B_h^2}{8\pi} > \tfrac{1}{2}\frac{\gamma {\mathcal{M}^{\prime}}^2}{1+T_c/T_h}P_{\rm ICM},
\end{equation}
where $B_c$ and $B_h$ are the magnetic field strengths in the cold and hot sides of the contact edge respectively, and $\mathcal{M}^{\prime}$ is the Mach number of the shear flow. In the case of NGC~1404, we take $\mathcal{M}^{\prime}= 1.1\sin\theta^{\prime}=0.6$; values of other parameters are chosen for Region ISM and Region ICM.
Conversely, the presence of KHI in NGC~1404 suggests that its magnetic field strength is no stronger than 
$\sqrt{4\pi P_{\rm mag}}=5\mu$G. 
Note that this upper limit is for an ordered magnetic field. If the field is tangled KHI can develop in the presence of substantially stronger magnetic fields.

\subsection{\bf Magnetic draping}

\begin{figure}[h]
   \centering
       \includegraphics[width=0.5\textwidth]{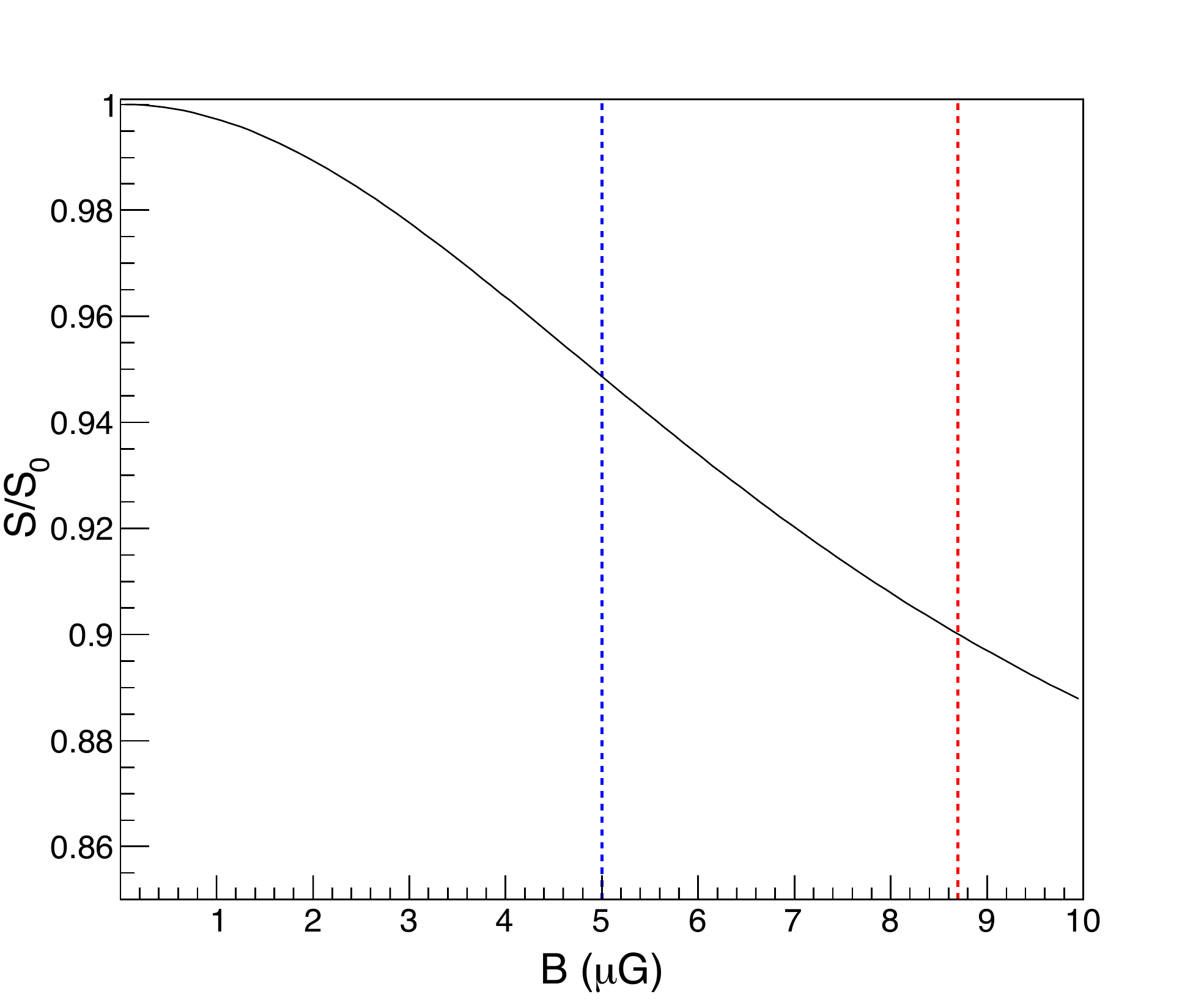}
        \includegraphics[width=0.5\textwidth]{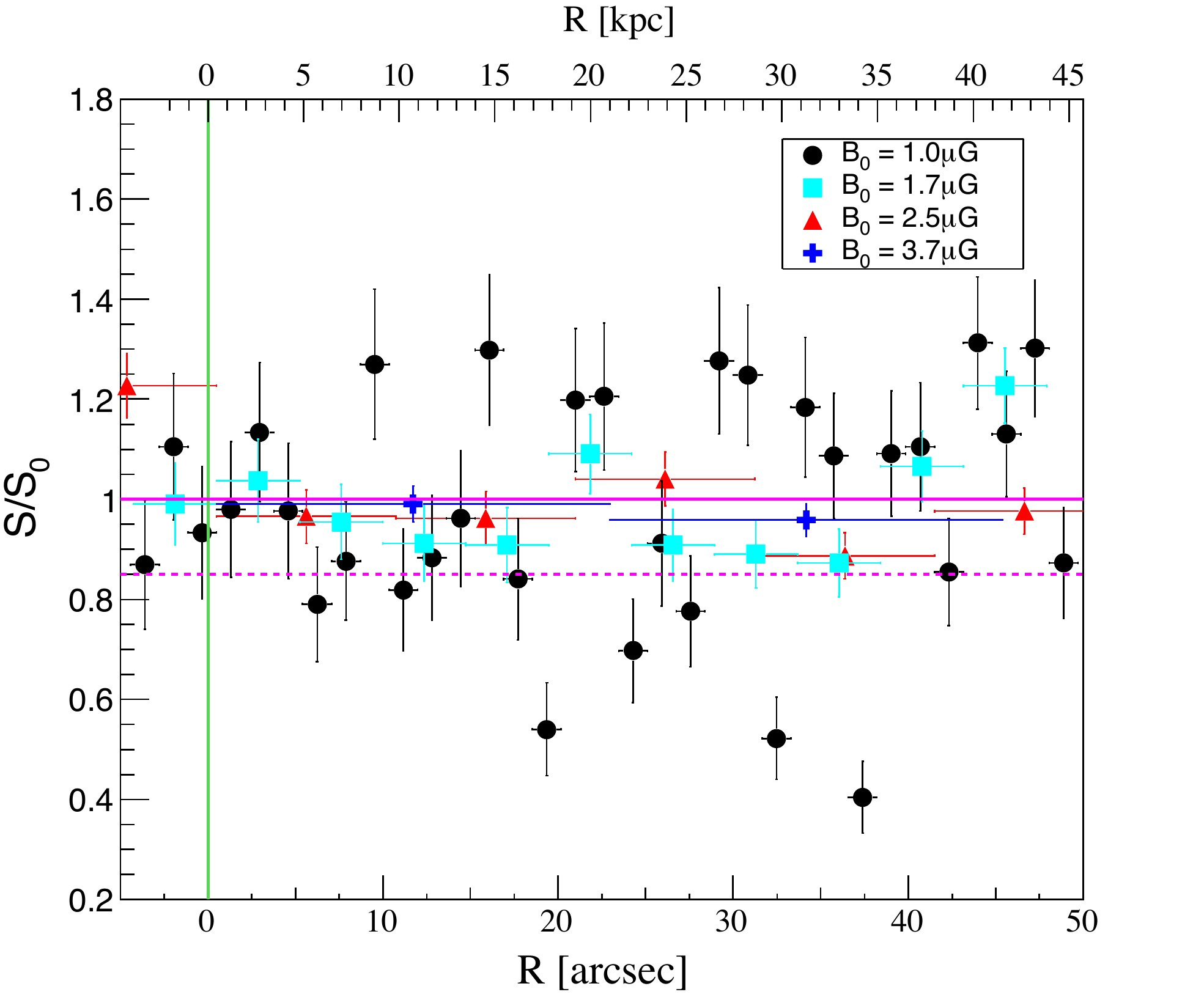}
%\vspace{-0.3cm}
\figcaption{\label{fig:layer}. {\it top:} the expected surface brightness of the magnetic draping layer relative to that of the free stream as a function of the field strength in the case of NGC~1404. Blue dashed line: the upper limit of field strength based on the pressure of KHI. Red dashed line: magnetic pressure equals thermal pressure ($\beta=1$). {\it bottom:} solid symbols: observed surface brightness profiles in the 1.3--4.0 keV energy band in front of the leading edge divided by the ICM $\beta-$profile. 
Green line indicates the location of the leading edge. Magenta lines mark the range of the y-axis in the top figure.}
\end{figure} 

As discussed in the previous section, magnetic field in the ICM
can be swept up and stretched over the leading edge of the infalling
spheroid, creating a ``magnetic draping layer" of enhanced field
(Dursi \& Pfrommer 2008; Ruszkowski et al.\ 2014; Lyutikov 2006). The field strength in this layer is amplified relative to the
free stream and may be sufficient to expel the gas, causing a
reduction in the thermal pressure.
Consequently, we may observe a ``dip" in the surface brightness just outside the leading edge. 
Lyutikov (2006) gives the expected thickness of the draping layer as $L\sim D/\mathcal{M_A}^2$, where  $\mathcal{M_A}=\sqrt{\gamma\beta_0/2}\mathcal{M}$ is the Alfven Mach number and $\beta_0$ the ratio of the ICM thermal pressure and the free stream magnetic pressure. 
The Faraday rotation measurement of Fornax A (at the outskirt of the Fornax Cluster) gives a line-of-sight magnetic field of $B_{\parallel}=0.3$--1.3\,$\mu$G (Bland-Hawthorn et al.\ 1995).
As a generic estimate, we assume $B_0=1\mu$G so that $\beta_0={P_{\rm ICM}}/{\frac{{B_0}^2}{8\pi}}$=75 for the ICM of the Fornax Cluster. For a $\mathcal{M}=1.32$ flow, $L$ would be 150 pc (1.64$^{\prime\prime}$). In principle, a surface brightness variation on this length scale can be resolved by {\sl Chandra} observations. Lyutikov (2006) also gives the expected density in the draping layer $\rho\propto(\beta/(1+\beta))^{1/\gamma}$, where $\beta$ is determined by the field strength in the draping layer $\beta={P_{\rm ICM}}/{\frac{{B}^2}{8\pi}}$. 
The volume emissivity
varies as $\epsilon \propto \rho^2 \propto [\beta / (1 + \beta)]^{2 / \gamma}$, so
that the volume emissivity of the draping layer is reduced relative
to the free stream by the factor $\epsilon / \epsilon_0 = [\beta / (1
+ \beta)]^{2 / \gamma}$.  However, the ICM in front of NGC~1404 takes up less than 20\% of the total ICM emission along the line-of-sight, so that
the decrease in surface brightness is diluted to 
$S/S_0=0.177 (\beta/(1+\beta))^{2/\gamma}+0.823$. In Figure~\ref{fig:layer}-top, we show $S/S_0$ as a function of $B$. The blue dashed line marks an upper limit of $B=5\mu$G that we obtained in \S4.5.  In Figure~\ref{fig:layer}-bottom, we plot the observed surface brightness profile in front of the leading edge divided by the best-fit ICM $\beta$-profile; we restricted this analysis to the 1.3-4.0 keV energy band to minimize the ``contamination" of the ``eddies". The range of the ICM surface brightness fluctuation exceeds the expected decrease; thus we are unable to put constraints on magnetic draping for $B_0=1.0\mu$G. 
We experimented with larger values of $B_0$ (i.e., $B_0=1.7\mu$G, $B_0=2.5\mu$G, and $B_0=3.7\mu$G) (Figure~\ref{fig:layer}-bottom). The range of the ICM fluctuation is reduced. Still, no significant deficit in the surface brightness just outside the contact edge is detected. Conversely, we may put an upper limit of $4\mu$G on the free stream magnetic field based on the absence of a detectable draping layer for $B_0=3.7\mu$G.

\section {\bf Conclusions}
We present the results of a very deep {\sl Chandra} observation of NGC~1404, a nearby gas rich galaxy falling through the intracluster medium of the Fornax Cluster. 
We investigated various transport processes occurring between the interstellar medium and the intracluster medium. 
We summarize our main results as follows:

$\bullet$ We put an upper limit of 5\% Spitzer on the viscosity of the hot cluster plasma based on the presence of eddies generated by Kelvin-Helmholtz instability. 

$\bullet$ We observe mixing between the cold ISM and hot ICM in its stripped tail, also indicative of a low viscosity plasma. 

$\bullet$ Across the leading edge we observe a jump in density over a scale
smaller than the mean free path, suggestive of suppressed electron diffusion. 

$\bullet$ The magnetic field in the plasma must be non-zero to suppress the diffusion, but an
ordered magnetic field cannot exceed $5\mu$G to allow Kelvin-Helmholtz instability to develop.

$\bullet$ We seek to find evidence for magnetic draping, a surface brightness deficit just outside the contact edge. The absence of this deficit on a scale of $20^{\prime\prime}$ allows us to put an upper limit of $4\mu$G on the ICM magnetic field.

%% References section
\section{\bf Acknowledgments}
We acknowledge helpful discussions with John ZuHone and Mateusz Ruszkowski.  
This work was supported by {\sl Chandra} Awards GO1-12160X and GO2-13125X issued by the
{\sl Chandra} X-ray Observatory Center which is operated by the Smithsonian Astrophysical Observatory under NASA contract NAS8-03060.

\end{document}